\documentclass{article}
\usepackage{graphicx}
\usepackage[square,numbers]{natbib}
\usepackage{hyperref}
\usepackage{lineno,hyperref}
\usepackage[none]{hyphenat}
\usepackage{doi}

\newcommand{\ave}[1]{\langle #1\rangle}

\begin{document}

\begin{flushleft}
{\Large
\textbf{Non-systemic transmission of tick-borne diseases: a network approach.}
}



Luca Ferreri$^{1,2,\ast}$, Paolo Bajardi$^{1,2}$, Mario Giacobini$^{1,2,3}$\\
\bf{1} Computational Epidemiology Group, Department of Veterinary Sciences, University of Torino, Italy
\\
\bf{2} Applied Research on Computational Complex Systems Group, Department of Computer Science, University of Torino, Italy
\\
\bf{3} CSU-Complex Systems Unit, Molecular Biotechnology Centre, University of Torino, Italy
\\
$\ast$ E-mail: luca.ferreri@unito.it
\end{flushleft}
\section*{Abstract}
\textrm{Tick-Borne diseases can be transmitted via non-systemic (NS) transmission. This occurs when tick gets the infection by co-feeding with infected ticks on the same host resulting in a direct pathogen transmission between the vectors, without infecting the host. This transmission is peculiar, as it does not require any systemic infection of the host. The NS transmission is the main efficient transmission for the persistence of the Tick-Borne Encephalitis virus in nature. By describing the heterogeneous ticks aggregation on hosts through a \hyphenation{dynamical} bipartite graphs representation, we are able to mathematically define the NS transmission and to depict the epidemiological conditions for the pathogen persistence. Despite the fact that the underlying network is largely fragmented, analytical and computational results show that the larger is the variability of the aggregation, and the easier is for the pathogen to persist in the population.}



\section{Introduction}
	Infections have always affected animals and humans, and, in the last centuries, an increasing research activity has been devoted in understanding infective processes and in depicting viable containment strategies. Since the pioneering work of Bernoulli in the eighteenth century \cite{Bernoulli1760} and more recently of Kermack and McKendrick \cite{Kermack}, mathematical models have played an important role in this area, helping field operators and physician to reveal important aspects of the infection dynamics or to test and validate health policies. In addition, in last decades, thanks to the increasing availability of computing power, the field has grown in interest and results also thanks to the introduction of new mathematical and computational frameworks. Cellular Automata, \cite{Wolfram}, Agent Based Models, e.g. \cite{Chowell2003,Eubank2004}, and more in general simulation approaches are becaming essential tools in epidemiology by supporting theoretical and analytical results or even revealing gaps of knowledge.
	
	In this context, recently some seminal works \cite{Becker1990,Ball1997,pastor-satorras2001,newman2002,May2001}  unveiled the important role of the contact structure of the epidemiological units on epidemic spreading, demonstrating that the heterogeneity of the number of  contacts between individuals substantially increases the probability of pathogens invasion. At the same time, many studies on real-world data showed the pervasiveness of  heterogeneous contact structures in many social \cite{Martin2006,Apicella2012,Miritello2013}, biological \cite{Jeong2001,Dunne2002,Dunne2002_b,Giot2003,Bascompte2003,Caretta2007,Mersch2013,Häuser2014}, technological, and infrastructural \cite{Latora2002,Sen2003,Pastor-satorras2004,Barrat2004,Kalapala2006} networks. Such heterogeneity implies that the distribution of the number of contacts per individual is fat tailed (thus following distributions like lognormal, Weibull, Pareto, etc.). Therefore, we should carefully consider an homogeneous mixing hypothesis (i.e. individuals interacting homogeneously with each other) when modelling an epidemiological scenario, and in many cases methods in which the heterogeneity is fully taken into account should be preferred. For further details, the reader can consult to the main references for social networks analysis \cite{wasserman,scott} and for networks modeling \cite{newman-book,Dyn-Proc-Nets,caldarelli}.

	Lately, some works explored dynamical processes occurring on networks evolving in time \cite{Isella2011,Miritello2011, Karsai2011, Rocha2013, Ferreri2014, Iribarren2009, Vazquez2007, Eames2004, Perra2012, Liu2014,Volz2009, Gross2006, Taylor2012,Massaro,Granell,Valdano2014,Valdano2015}. In particular, Perra and collagues \cite{Perra2012,Liu2014} proposed an activity driven network model where at each time-step edges are ''fired'' from nodes accordingly their potential. Results show that by distributing heterogeneously the activity potential to nodes the model is able to reproduce graphs with skewed degree distributions similar to those observed in reality. Furthermore, Perra's group showed that the an epidemic process occurring on networks deriving from such model depends on the activity distribution of nodes and in particular to its  heterogeneity. More recently, Valdano and colleagues \cite{Valdano2014}, proposed a methodology that given a temporal network is able to detect the epidemic threshold with a very good accuracy. In particular, the  authors described the epidemic threshold in terms of the spectral radius of a matrix defined by the network topology and the disease features.
	
	In this manuscript, we are interested in a challenging epidemiological problem: to model the spreading of a tick-borne disease (TBD) occurring via non-systemic (NS) transmission. TBDs are diseases that are naturally maintained in a complex cycle of vectors (ticks) and those hosts on which ticks take their blood meals (usually mammalian, reptiles, and birds according to ticks habits). TBDs differ from other vector-borne diseases (such as malaria, the most known mosquito-borne disease) mainly thanks to ticks' peculiarities:  ticks have a limited mobility and usually wait for a host instead of seeking for it, and in many cases they make a single complete blood meal on a single host before moulting. 

	A TBD can be usually maintained in nature by three transmission routes. First, the systemic transmission, occurring when a tick (respectively a host) gets the infection from a infective host (respectively a tick) during a blood meal. Then, the transovarial transmission, occurring when the pathogen is transmitted from a infected female tick to its offspring. Finally, the non systemic (NS) transmission, occurring when a tick gets the infection by co-feeding with infected ticks on the same host. The peculiarity of the NS transmission is that it does not require any systemic infection of the host. As an example of a pathogen exploiting these three transmission routes, the reader can refer to Lyme borreliosis, whose ethiological agent are bacteria of the \textit{Borrelia burgdorferi} sensu lato compelx \cite{gernhumair1998,Mannelli2012}. 

In this article, we are interested in the Tick-Borne Encephalitis (TBE) which is an emerging TBD in Europe \cite{RandolphSumilo2007}, as it causes the most important arboviral infection of the human central nervous system in Eurasia, resulting in long-term sequelae and, in some cases, to death \cite{Donoso}. TBE virus, in Europe, is maintained in a cycle involving as vector \textit{Ixodes ricinus} ticks. \textit{I. ricinus} is a three stages (larva, nymph, and adult) hard-tick which needs a complete blood meal on a single host before moulting from a stage to the following. Rodents are the main host for these ticks at juvenile stages, larval and nymphal, while adult ticks are found principally on other larger mammalians, e.g. deer \cite{Lindquist2008}. However, since rodents are indicated as the main reservoir for TBE virus \cite{Mansfield} and the transovarial transmission seems to be negligible for its persistence \cite{Nuttall2003}, the most important stages in the epidemiological cycle of this pathogen are the larval and nymphal ones. In addition, TBE virus causes a very short systemic infection in rodents, \cite{Randolph1996}. Consequently, several studies \cite{Rosa2003,Hartemink2008,Harrison2012}  suggested that the NS transmission is the main efficient transmission for the persistence of the TBE virus in nature. 

	In this epidemiological scenario, the burden of ticks on rodents and the consequent contact pattern of ticks co-feeding on rodents, become the key factors when modelling this pathogens' dynamics. Tick aggregations on rodents have been recognized as highly heterogeneous \cite{Randolph1975,Shaw1995,Shaw1998,Brunner2008,Bisanzio2010,Kiffner2011,Harrison2012,Ferreri2014a} with a large number of rodents parasitised by few ticks and a smaller number of rodents having a large number of ticks. Therefore, the homogeneous mixing approximation should be avoided when describing the contact structure between ticks and rodents. Accordingly, in the following, we are going to introduce a discrete-time mathematical model describing the NS transmission and the heterogeneous pattern of contact between ticks and mice. In this model, we are able to detect a threshold condition for the epidemic persistence. We further investigate this threshold by microscopical simulations of the epidemic spreading.

\section{Model}
	
	In order to model the heterogeneous contact pattern between ticks and hosts, we use a particular class of graph, dynamical star graphs $S_k$. An $S_k$ is defined by a central node (the mouse) and $k$ peripheral nodes (the ticks) connected with the central node by edges called rays. No connection is allowed between peripheral nodes, resulting in a particular type of bipartite graph.

We further define a constellation $\mathcal{S}(N,p)$ as the set of $N$ stars $S_{k_i}$ with each $k_i$ ($i=1,\ldots,N$) sampled from a fixed probability distribution $p(k)$. Given the constellation $\mathcal{S}(N,p)$, the number of peripheral nodes is $M=N\sum_k kp(k)=N\ave{k}$. We partition them according to their epidemic status (Susceptible, $S$, and Infective, $I$), while central nodes are involved in the epidemic process only as bridges between peripheral nodes. 
		
	Over this structure, to model the pathogen spreading process, we define a simple Susceptible-Infected-Susceptible (SIS) model \cite{Anderson&May,rohani,brauer}, as already proposed for TBD modeling in \cite{Bisanzio2010}. In this framework, let denote $\pi(t)$ the prevalence at time $t$ (i.e. the fraction of infective nodes), and $\beta$ and $\mu$ the infection and recovery probabilities, respectively. 

At each discrete time step $t$, let consider an instance of the constellation $\mathcal{S}(N,p)$. Consequently, the probability that a peripheral node is on a star $S_h$ is $\frac{h p(h)}{\ave{k}}$. Let us consider a peripheral node connected to a central node together with other $h-1$ other nodes. The expected number of infective nodes in the neighborhhod of the central node is $\pi(t)\left(h-1\right)$. Therefore, supposing that the considered node is susceptible, the probability that it gets the infection is
	\[1-\left(1-\beta\right)^{\left(h-1\right)\pi(t)}.\]
	As consequence, the probability that a susceptible node at time $t$ is infective at time $t+1$ is
	\[\gamma(t)=\sum_h\left(\left(1-\left(1-\beta\right)^{(h-1)\pi(t)}\right)\frac{h}{\ave{k}}p(h)\right),\]
	and the evolution of the prevalence in time is described by the following discrete-time Markov-chain
	\begin{equation}\pi(t+1)=F(\pi(t))=(1-\mu) \pi(t)+\left(1-\pi(t)\right)\gamma(t)+\mu\pi(t)\gamma(t)\label{evo}\end{equation}
	where the first term of $F(\pi(t))$ is the probability that a infective node at time $t$ is not recovered at time $t+1$, the second term is the probability that a susceptible node gets the infection, and the third term is the probability that a infective node at time $t$ gets recovered but immediately it is reinfected \cite{gomez}.
	
\begin{figure}[ht!]
	\centering
		\includegraphics{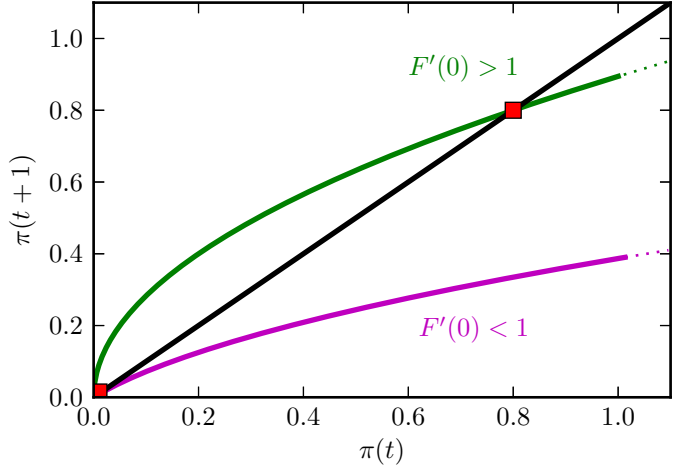}
		\caption{An illustrative picture of the solutions of eq.~(\ref{evo}). When $F'(0)>1$ the non-trivial steady state, $\hat{\pi}$, exists while when $F'(0)<1$ the only equilibrium point is the origin.}\label{fig-conditio}
	\end{figure}

	Now, we analyze the phase diagram of this process by studying the steady state of eq.~(\ref{evo}) in the bound interval $[0,1]$. The trivial solution $\tilde\pi=0$ is a steady point. Moreover, since $F(1)\leq1$, $F'(\pi)>0$, and $F''(\pi)<0$ there exists one and only one non trivial equilibrium point, $\hat{\pi}$, if and only if $F'(0)>1$. We sketch in Fig.~\ref{fig-conditio} the two different cases. In conclusion, the non trivial equilibrium exists if and only if:
	\begin{equation}\frac{-\log(1-\beta)}{\mu}>\frac{\ave{k}}{\ave{k^2}-\ave{k}},\label{conditio}\end{equation}
	where $\ave{k^2}$ is the second moment of $p(k)$.
	
	Condition (\ref{conditio}) is of great interest for at least two reasons. On one hand, it demonstrates the presence of clear phase change in the epidemic behavior: for large enough $\beta$, i.e. for $\beta>\beta_c$, the pathogen reaches the condition for its persistence. On the other hand it binds the critical infection probability, $\beta_c$, to the heterogeneity of the distribution of rays in constellations. This means that the larger the heterogeneity, the smaller the $\beta$ is needed by the pathogen to mantain its persistence. In addition, we could also state that when only the trivial equilibrium point exists, since $\left|F'(0)\right|<1$ is hold, the zero solution is asymptotically stable. On the other hand, for $F'(0)>1$ we have that the the trivial equilibrium is unstable and $\hat{\pi}$ is asymptotically stable \cite{Elaydi}.


\section{Numerical Results and Analysis}
	To show the validity of the analytic approach presented above, we perform both numerical integrations of eq.~\ref{evo} and Monte Carlo (MC) simulations of the epidemic process. In particular, since we are interested in depicting the influence on the epidemic threshold of the aggregation of ticks on hosts, we need a network model that, while keeping constant the total number of edges in the structure, is able to tune the rays heterogeneity. 

To this end, we first generate a constellation $\mathcal{S}(N,p)$, in which $p$ is a discrete Pareto distribution defined as $p(k)\sim k^{-(\gamma+1)}$ for $k\geq k_m$ with the scale parameter $k_m=2$ and the shape parameter $\gamma=1.25$ (arbitrarly chosen to have an highly heterogeneous distribution). Then for a given $\alpha \in [0,1]$, we randomly reassign the fraction $\alpha$ of the peripheral nodes to central nodes. We further impose that a ray departing from a central node of degree two (the minimum degree since $k_m=2$) could not be reallocated to avoid central nodes with a number of edges less than $k_m$. We call the so obtained constellation $G(\mathcal{S}(N,p),\alpha)$. 

We further remark that for a given $G(\mathcal{S}(N,p),\alpha)$, there exists a probability distribution $q$ for which $\sum_k kp(k)=\sum_h hq(h)$ and $\sum_k k^2p(k)\geq\sum_h h^2q(h)$, and such that $G(\mathcal{S}(N,p),\alpha)=\mathcal{T}(N,q)$. In addition, we would also highlight that given a constellation $\mathcal{S}(N,p)$ the graphs $G(\mathcal{S}(N,p),\alpha_0)$ and $G(\mathcal{S}(N,p),\alpha_1)$ have the same number of edges but differ in their degree heterogeneity which is driven by $\alpha$ (thus ranging from the one induced by the Pareto distribution, $\alpha=0$, to the one of a random graph, $\alpha=1$). 

\begin{figure}[ht!]
	\includegraphics{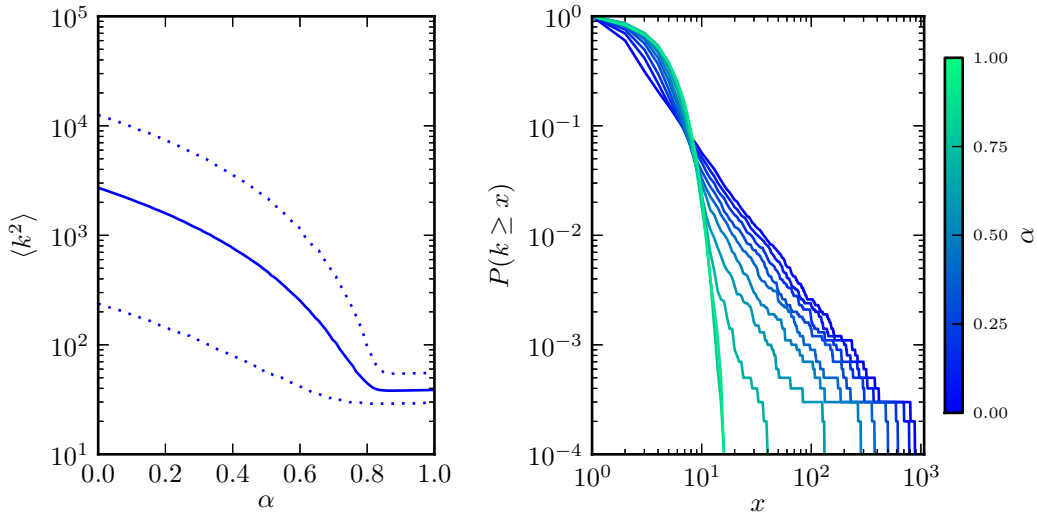}
	\caption{Left: mean and $95\%$ confidence interval of heterogeneity in $G(\mathcal{S}_i(N,p),\alpha)$, as measured by $\ave{k^2}$, with $i=1,\ldots,100$. Right: degree distributions for different values of $\alpha$ for a graph $G(\mathcal{S}(N,p),\alpha)$. For both plots $N=10^4$.}\label{distr}
	\end{figure}

In this framework, the constellations $G(\mathcal{S}_0(N,p),\alpha)$ and $G(\mathcal{S}_1(N,p),\alpha)$ are different in the sense that they are generated from two different, $\mathcal{S}_0(N,p)$ and $\mathcal{S}_1(N,p)$, initial constellations. Now, the heterogeneity of a constellation generated from a Pareto distribution, $\mathcal{S}_0(N,p)$ could be very different if compared to $\mathcal{S}_1(N,p)$. For instance, we observe that $\ave{k^2}$ could vary of two order of magnitude (e.g. right panel of Figure~\ref{distr}). 

Therefore, we want to remark that the direct comparison of $\beta_c$ observed for $G(\mathcal{S}_0(N,p),\alpha_0)$ and for $G(\mathcal{S}_1(N,p),\alpha_1)$ could lead to misleading results. For instance, supposing $\alpha_0<\alpha_1$, from the analytic results we could expect the lowest critical infection probability associated with $\alpha_0$. But, in the case that $\ave{k^2}_{\mathcal{S}_0(N,p)}<\ave{k^2}_{\mathcal{S}_1(N,p)}$, it could happen that the lowest epidemic threshold is associated to $\alpha_1$. 

For a graphical representation of the graph heterogeneity changing as a function of $\alpha$, we plot on the right panel of Fig.~\ref{distr} the degree distributions of a graph $G(\mathcal{S}(N,p),\alpha)$, while on the left panel we plot the induced second moment $\ave{k^2}_{G(\mathcal{S}_j(N,p),\alpha)}$ for $j=1,\ldots,100$ as function of $\alpha$.
	
\begin{figure}[ht!]
	\centering
	\includegraphics{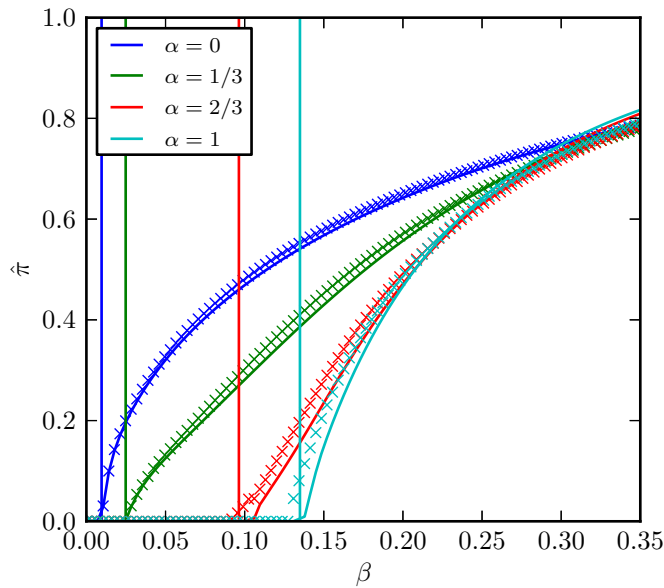}
	\caption{Stable prevalence, $\hat{\pi}$, as function of $\beta$. $N=10^4$, $\mu=0.75$. Dotted lines represent average of prevalence at reached equilibrium of MC simulations in non-trivial equilibria, full lines those of numerical integration. Vertical lines are the ET as estimated by Eq.~\ref{conditio}.}\label{curve_epi}
	\end{figure}

	Once fixed $N=10^4$, we generate a constellation $\mathcal{S^*}(N,p)$ and given an initial condition, $\pi(t_0)=0.01$, we recursively iterate Equation~(\ref{evo}) on $G(\mathcal{S^*}(N,p),\alpha)$ for different values of $\alpha$. After few iterations (less than $100$) the equilibrium point (trivial or not, according to the epidemic parameters) is reached. Thus, we plot in Figure~\ref{curve_epi} the obtained stable point in function of the epidemic parameters, $\beta$ and $\mu$ and of $\alpha$. The epidemic threshold predicted by analytic arguments and the numerical simulations are in good agreement.

	In Figure~\ref{curve_epi} we also plot the average prevalence of MC simulations where the steady non-trivial equilibrium is reached. In particular, on $G(\mathcal{S^*}(N,p),\alpha)$, for different combinations of parameters $(\beta,\mu,\alpha)$, we start $300$ simulations from one infected peripheral-node (randomly chosen). Then, at each iterations $t$ the peripheral nodes with their health status are permuted around central nodes under the constrain given by $G(\mathcal{S^*}(N,p),\alpha)$ and the NS transmission between infectious and susceptible peripheral nodes occurs with probability $\beta$. Afterwards, the recovery process could occur with probability $\mu$. Results of MC simulations give further evidences of the accurate detection of the critical transmission probability. 

\begin{figure}[ht!]
	\centering
	\includegraphics{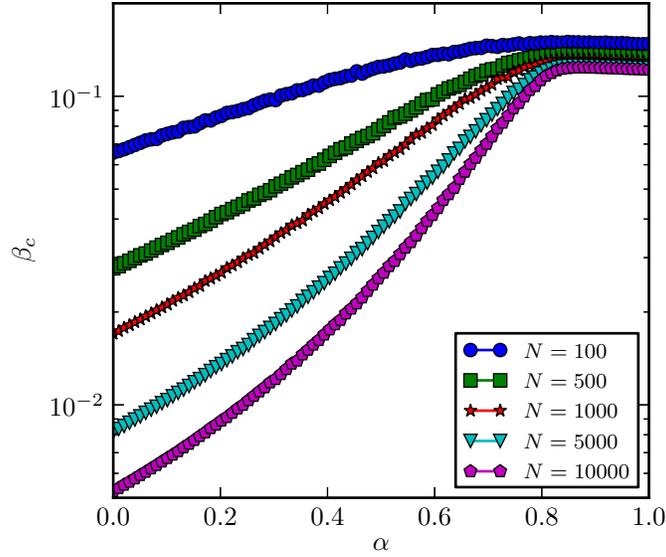}
	\caption{Averaged critical infection parameter as function of $\alpha$ for different network sizes $N$}\label{vanish}
	\end{figure}

	We further investigate the critical transmission parameters, $\beta_c$, of $G(\mathcal{S}_j(N,p),\alpha)$ with $j=1,\ldots,100$ and in particular their dependence on the network size $N$. To this end, for each triplets $(i,\alpha,N)$ we replicate $100$ times the epidemic threshold detection. We plot the average values in Figure~\ref{vanish}.
	
As expected, we found that $\beta_c$ is an increasing function of $\alpha$ and decreasing of $N$. In particular, we have that the $\beta_c$ in $\alpha=0$ starts from a point depending on the size $N$ (the larger $N$, the larger the heterogeneity and thus the lower the critical transmission probability). Then, $\beta_c$ increases driven by the reshuffling parameter, $\alpha$, reaching a plateau which is almost the same for the different $N$.


\section{Discussion and Conclusion}
	In this work, inspired by the peculiar features of the TBE virus transmission, we propose a model of epidemic spreading on dynamic star graphs. At each time step a constellation of star graphs is generated and the infection is spread among rays of the stars. This allows us to reproduce the most important aspects of the TBE dynamics for its persistence: the high aggregation of ticks on hosts and the NS transmission. Analytic results, validated by numerical investigations, suggesting that the epidemic threshold $\beta_c$ depends to the second moment of the distribution of the number of rays of star graphs. This is a well known result for the epidemic dynamics on networks \cite{pastor-satorras2001,May2001,newman2002}. However, it is somehow surprising that also in the case of a graph made of disconnected star-like structures  holds the very same description of the invasion threshold. Being the nodes (i.e. ticks) randomly re-attached at every time-step, the system under investigation resembles a spreading process on annealed networks. This result has important consequences for the epidemiology of TBE: in fact, it suggests that the aggregation of ticks on host is a favorable condition for the persistence of the viruses. In this way, the correct identification (and modelization) of ticks burden on mice is the key-factor for the understanding of the tick-borne diseases.
	
	It is important to stress out that recent literature is rich of models defined for temporal networks (e.g.  \cite{Isella2011,Miritello2011, Karsai2011, Rocha2013, Ferreri2014, Iribarren2009, Vazquez2007, Eames2004, Perra2012, Liu2014,Volz2009, Gross2006, Taylor2012,Valdano2014,Valdano2015}). Unfortunately, such generative models are not suitable to describe the NS transmission of TBEv where the underlying network structure is intrisically memory-less being vectors hosted for one blood meal only before developing in the new stage and attaching a new host. We therefore defined our dynamical model made of unconnected star-like graphs where nodes are fully reshuffled after each time-step to fully describe the epidemiological peculiarities of the process under study. However, a relevant aspect of our model is that it is general enough to be extended to other systems. For instance, diseases spreading occurring among people using transport means could be described by this model. In this scenario, transport means are the central nodes while passengers are the peripheral nodes among which the infection are spread. However, an extensive research in this direction will be subject of future work.
	
	Moreover, other future works will be devoted in a better integration of this model to the ecological complexity of the spreading of TBE virus. In particular, we would like to take into account the seasonality of ticks and rodents behavior and to explicit the ticks stages. An other interesting extension would be to apply our model to other tick-borne diseases and thus to integrate it with more complex aspects of their dynamics.

\section*{Acknowledgements}
LF acknowledges the \href{http://www.progettolagrange.it/en}{Lagrange Project on Complex Systems} -- CRT and ISI Foundation. MG acknowledges local funding of the University of Torino.

\bibliography{biblio}

\end{document}